\documentclass[10pt,onecolumn]{article}
\usepackage{joss}

\usepackage[style=apa, backend=biber]{biblatex}
\jossRepoURL{https://github.com/dogusariturk/HEACalculator}
\jossArchiveDOI{10.5281/zenodo.3590318}
\jossDocsURL{https://dogusariturk.github.io/HEACalculator/}
\jossDate{Jun 2026}
\jossLogo{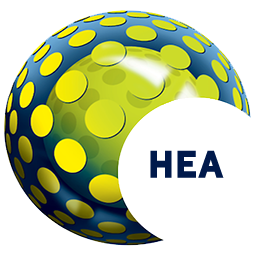}
\jossFooter{Sarıtürk, Kalay \& Arróyave (2026). \textit{HEACalculator}: An Open-Source Python Tool for Thermodynamic Property Calculation and Solid Solution Prediction in High-Entropy Alloys}

\jossauthor[1]{Doğuhan Sarıtürk}
\jossorcid{0000-0003-0008-1165}
\corrauthor{sariturk@tamu.edu}

\jossauthor[2]{Yunus Eren Kalay}
\jossorcid{0000-0002-5514-5202}

\jossauthor[1,3,4]{Raymundo Arróyave}
\jossorcid{0000-0001-7548-8686}

\affiliation[1]{Department of Materials Science and Engineering, Texas A\&M University, USA}
\affiliation[2]{Department of Metallurgical and Materials Engineering, Middle East Technical University, Turkey}
\affiliation[3]{J. Mike Walker '66 Department of Mechanical Engineering, Texas A\&M University, USA}
\affiliation[4]{Wm Michael Barnes '64 Department of Industrial and Systems Engineering, Texas A\&M University, USA}

\title{HEACalculator: An Open-Source Python Tool for Thermodynamic Property Calculation and Solid Solution Prediction in High-Entropy Alloys}

\begin{document}

\sloppy

\section{Summary}

High-entropy alloys (HEAs) have attracted sustained interest since their introduction by \textcite{cantor2004} and \textcite{yeh2004} because multi-principal-element compositions can exhibit unusual combinations of strength, thermal stability, and functional performance \parencite{george2019,miracle2017}. A recurring problem in HEA design is determining whether a candidate composition is likely to form a single-phase solid solution or instead separate into multiple phases or intermetallic compounds. That question sits early in the alloy-design workflow because it shapes which compositions require further thermodynamic analysis, synthesis, and experimental validation.

HEACalculator is an open-source Python package for calculating thermodynamic and structural descriptors used in HEA research and for evaluating published solid-solution formation rules in a single place. It computes sixteen commonly used quantities, including mixing enthalpy, configurational entropy, valence electron concentration, Hume-Rothery electron-to-atom ratio, atomic size mismatch, electronegativity mismatch, and derived stability parameters such as Omega, Lambda, and Phi, and it evaluates eight published prediction criteria \parencite{yang2012,guo2011,wang2015,singh2014,ye2015scr,troparevsky2015,senkov2016,king2016}. The package combines a curated elemental and binary-interaction dataset with three access modes: a command-line interface (CLI), a desktop graphical user interface (GUI), and a Python application programming interface (API) for programmatic use in notebooks and screening workflows.

\section{Statement of Need}

Predicting solid-solution formation in HEAs remains a practical bottleneck in alloy design because the field relies on several phenomenological criteria distributed across separate papers, each with its own notation, parameter definitions, and validation set. Researchers who want to compare models or reuse them in automated studies must therefore reconstruct overlapping calculations, assemble supporting datasets, and track subtle differences in implementation. This makes even routine composition screening unnecessarily difficult and complicates reproducible comparison across studies.

HEACalculator addresses that gap by computing sixteen thermodynamic and structural descriptors (\autoref{tab:descriptors}) and providing a validated, reusable implementation of all eight published criteria in one package (\autoref{tab:models}). It combines a database of 118 elements, 2,628 binary mixing enthalpies from \textcite{takeuchi2005}, and 435 binary formation enthalpies derived from density functional theory (DFT) data \parencite{troparevsky2015}, together with utilities for single-composition analysis, parameter sweeps, and batch processing. The intended users are materials scientists who need a transparent tool for exploratory alloy design, high-throughput screening, or preparation of experimentally testable candidate compositions.

\begin{table}[t]
\centering
\small
\caption{Thermodynamic and structural descriptors computed by HEACalculator for a candidate alloy composition. The sixteen quantities cover mixing thermodynamics, atomic size and electronegativity mismatch, electron-structure descriptors, and derived stability parameters used in published solid-solution formation criteria. Here $c_i$, $w_i$, and $V_i$ are the atomic fraction, atomic weight, and atomic volume of element $i$; $\bar{r}$ and $\bar{\chi}$ are composition-weighted averages; $R$ is the gas constant; $H_a$ is the binary mixing enthalpy (J/mol); $S_E$ is the hard-sphere excess entropy averaged over BCC and FCC packing fractions; and $(e/a)_i$ counts only outer s+p electrons (d and f electrons excluded), following the Hume-Rothery convention.\label{tab:descriptors}}
\renewcommand{\arraystretch}{1.5}
\begin{tabularx}{\linewidth}{llXl}
\toprule
\textbf{Symbol} & \textbf{Parameter} & \textbf{Formula} & \textbf{Unit} \\
\midrule
$\rho$ & Density & $\sum w_i c_i / \sum V_i c_i$ & g/cm$^3$ \\
$T_m$ & Melting temperature & $\sum c_i T_{m,i}$ & K \\
$\Delta H_\text{mix}$ & Mixing enthalpy & $4\sum_{i<j} c_i c_j H_{ij}^\text{mix}$ & kJ/mol \\
$\Delta S_\text{mix}$ & Mixing entropy & $-R\sum c_i \ln c_i$ & J/(K$\cdot$mol) \\
$\Delta H_f$ & Formation enthalpy & $4\sum_{i<j} c_i c_j H_{ij}^f$ & meV/atom \\
$\min(\Delta H_f)$ & Min.\ binary formation enthalpy & $\min_{i<j}(H_{ij}^f)$ & meV/atom \\
VEC & Valence electron concentration & $\sum c_i \text{VEC}_i$ & --- \\
$e/a$ & Hume-Rothery electron-to-atom ratio & $\sum c_i (e/a)_i$ (s+p electrons only) & --- \\
$\delta$ & Atomic size difference & $100\sqrt{\sum c_i(1-r_i/\bar{r})^2}$ & \% \\
$\delta_\text{CN12}$ & Atomic size difference (CN12) & same formula, CN12 radii & \% \\
$\Delta\chi_\text{Allen}$ & Electronegativity diff.\ (Allen) & $100\sqrt{\sum c_i(1-\chi_i/\bar{\chi})^2}$ & \% \\
$\Delta\chi_\text{Pauling}$ & Electronegativity diff.\ (Pauling) & same formula, Pauling scale & \% \\
$\Omega$ & Omega & $T_m \Delta S_\text{mix} / |\Delta H_\text{mix}|$ & --- \\
$\gamma$ & Gamma & $\Omega_S / \Omega_L$ (solid-angle ratio) & --- \\
$\lambda$ & Lambda & $\Delta S_\text{mix} / \delta_\text{CN12}^2$ & --- \\
$\varphi$ & Phi & $(\Delta S_\text{mix} - |H_a|/T_m) / |S_E|$ & --- \\
\bottomrule
\end{tabularx}
\end{table}

\begin{table}[ht]
\centering
\small
\caption{Published criteria for solid-solution formation implemented in HEACalculator. The crystal-structure row predicts the expected phase (FCC, BCC, or HCP) from the valence electron concentration; Models~1--8 each classify a composition as \textit{Solid Solution}, \textit{Intermetallic}, or \textit{Multiple Phases} based on the stated threshold. All criteria are evaluated from the same set of cached thermodynamic descriptors, ensuring consistent classification regardless of the access mode used. Model~7 parameters $k_2$ and $T_\text{ann}$ are user-configurable, with defaults $k_2 = 0.6$ and $T_\text{ann} = T_\text{crit}$, where $T_\text{crit} = 0.55\,T_m$.\label{tab:models}}
\renewcommand{\arraystretch}{1.5}
\begin{tabularx}{\linewidth}{lXl}
\toprule
\textbf{Model} & \textbf{Solid-Solution Criterion} & \textbf{Reference} \\
\midrule
Crystal Structure & VEC $\geq 8$: FCC \newline VEC $\leq 6.87$: BCC \newline $2.5 \leq$ VEC $\leq 3.5$: HCP & \parencite{guo2011} \\
1 & $\Omega \geq 1.1$ \newline $\delta \leq 6.6\%$ & \parencite{yang2012} \\
2 & $-11.6 < \Delta H_\text{mix} < 3.2$\,kJ/mol \newline $\delta < 6.6\%$ & \parencite{guo2013} \\
3 & $\gamma < 1.175$ & \parencite{wang2015} \\
4 & $\lambda > 0.96$ \newline $0.24 \leq \lambda \leq 0.96$: multiple phases & \parencite{singh2014} \\
5 & $\varphi \geq 20$ & \parencite{ye2015scr} \\
6 & all binary $\Delta H_f \in (-T_\text{crit}\Delta S_\text{mix},\; 37)$\,meV/atom & \parencite{troparevsky2015} \\
7 & $k_1 < \Omega(T_\text{ann})(1-k_2)+1$; $k_2=0.6$, $T_\text{ann}=0.55\,T_m$ & \parencite{senkov2016} \\
8 & $F = \Delta G_\text{SS}/(-|\Delta G_\text{max}|) \geq 1$ & \parencite{king2016} \\
\bottomrule
\end{tabularx}
\end{table}

\section{State of the Field}

The closest direct peer is HEAPS, a free HEA screening program that computes a larger collection of semi-empirical descriptors and criteria, including mechanical-property heuristics, through a MATLAB-based graphical workflow \parencite{martin2022heaps}. HEAPS offers two operating modes: a single-composition calculator and an exploration mode capable of screening large numbers of candidate alloys simultaneously against user-defined composition and parameter constraints. HEAPS is therefore broader in screening scope than HEACalculator. However, the two projects make different research-software trade-offs and impose different runtime requirements. HEAPS runs exclusively on 64-bit Windows systems and requires either the free MATLAB Runtime for its standalone installer or an active MATLAB license to use the source function files directly. HEACalculator, by contrast, focuses more narrowly on the widely cited thermodynamic and structural descriptors used for solid-solution analysis, but packages them as a cross-platform, installable Python library with a documented API, command-line interface, and optional desktop GUI. That packaging makes the calculations directly reusable in notebooks, automated screening scripts, and testable pipelines without platform or proprietary-software constraints.

Within the Python ecosystem, general-purpose descriptor libraries such as matminer \parencite{ward2018matminer} and HEA-specific featurizers such as AutomaticFeaturizerMPEA \parencite{subedi2022} compute alloy properties as inputs for machine-learning workflows. Neural-network-based tools such as pyMPEALab \parencite{subedi2021pympealab} extend this further by coupling featurization to a trained classifier and presenting phase predictions through a graphical interface. These tools are designed for data-driven discovery pipelines rather than explicit criterion evaluation: their predictions emerge from learned model weights rather than named thresholds, so the correspondence between a classification outcome and any specific published criterion, reference dataset, or formula is not preserved. HEACalculator addresses a complementary need in the same Python ecosystem by implementing the formation criteria directly, producing predictions that are reproducible and traceable to individual source publications.

At the lighter-weight end of the ecosystem, model-specific scripts and spreadsheet calculators distributed as supplementary materials alongside individual publications are convenient for reproducing a particular set of results, but they are not designed as versioned, scriptable research dependencies. They generally provide less visibility into data provenance, software versioning, and automated regression testing than a packaged codebase intended for reuse. In that context, the scholarly contribution of HEACalculator is not that it is the broadest HEA design environment, but that it provides a transparent, Python-native, literature-validated implementation of eight published solid-solution criteria that can be inspected, cited, and embedded directly in reproducible computational materials workflows. Contributing to existing interactive tools would not remove the need for that programmatic, workflow-oriented capability.

\begin{figure}[h]
  \centering
  \includegraphics[width=\linewidth]{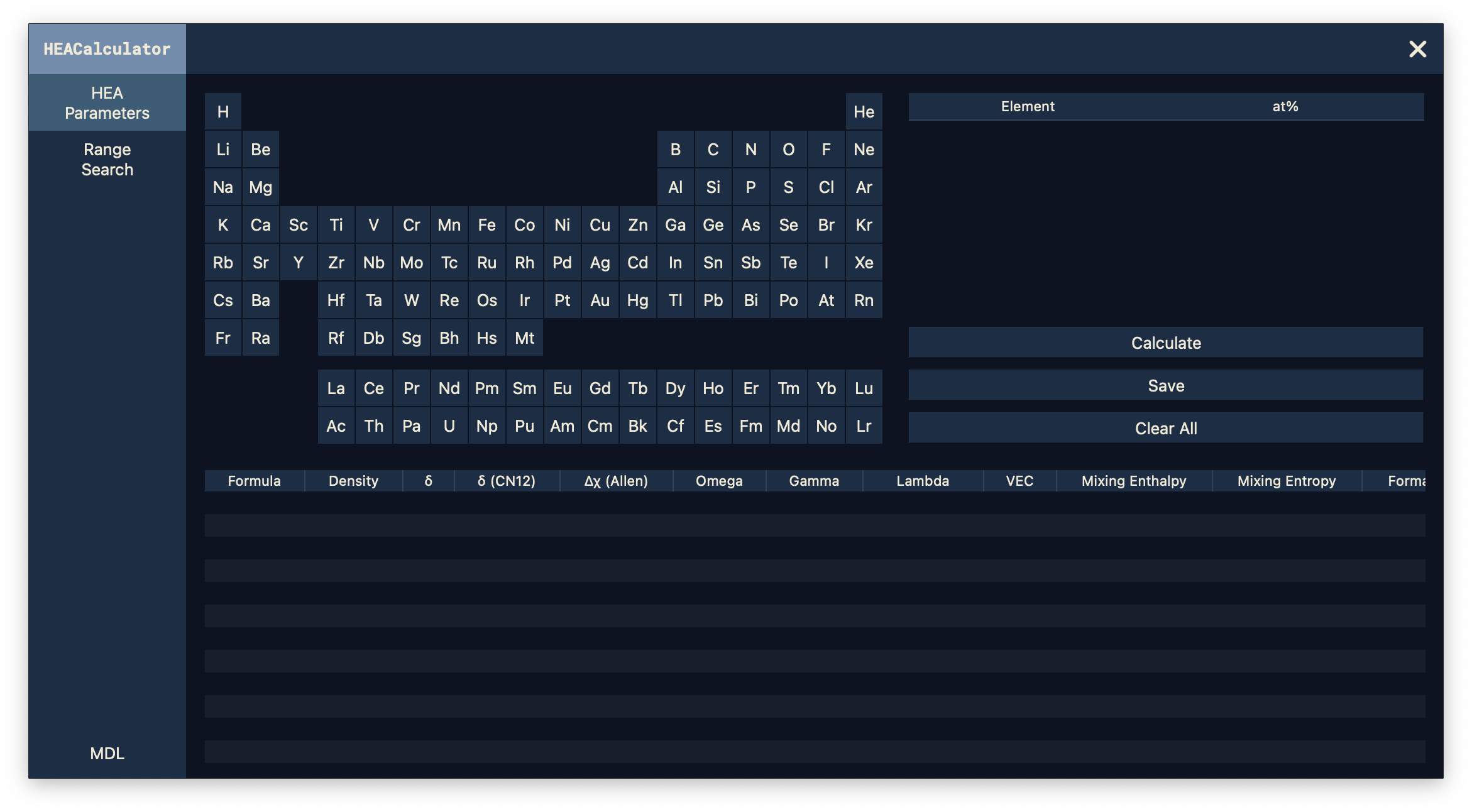}
  \caption{The HEACalculator graphical interface showing the HEA Parameters page. Users select elements from the periodic table, assign atomic-percent compositions, and calculate thermodynamic descriptors and solid-solution predictions in a single shared results view.\label{fig:figure1}}
\end{figure}

\section{Software Design}

HEACalculator is organized around a shared calculation core rather than around its user interfaces. Compositions supplied through the CLI, GUI, or Python API are first normalized by a common parser into an internal composition representation that supports compact formulas, explicit stoichiometries, and nested notation. That normalized state is then passed to a thermodynamic engine that computes the required descriptors once and reuses them across all downstream model evaluations. This separation keeps the interfaces thin and ensures that a composition produces the same result regardless of how the package is invoked.

The design emphasizes reproducibility and consistency in two ways. First, thermodynamic and structural quantities are evaluated lazily and cached per composition, which avoids duplicate arithmetic when several prediction models depend on the same intermediate values. Second, the prediction layer is kept separate from property calculation, so published model thresholds can be applied to a single validated set of descriptors rather than recomputed independently in each interface. This matters in HEA screening workflows, where small implementation differences can lead to inconsistent classification near model boundaries.

The package also relies on curated JSON-backed reference data for elemental properties, binary mixing enthalpies, and binary formation enthalpies. Keeping those datasets explicit inside the package makes the provenance of the calculations inspectable and simplifies reuse in automated studies. When required pair data are unavailable, typed exceptions are raised instead of silently substituting values, which makes incomplete coverage visible to calling code and helps users distinguish unsupported compositions from physically meaningful negative results.

This architecture was chosen to serve research use rather than only interactive exploration. The CLI supports single-alloy queries, composition-range sweeps, and CSV-based batch processing. The optional PyQt6 GUI supports desktop use (\autoref{fig:figure1}). The Python API supports direct integration into notebooks, scripts, and screening pipelines. Because all three interfaces share the same parser, datasets, and cached thermodynamic state, the package can function as both an end-user calculator and a reusable component in larger computational materials workflows.

\section{Research Impact Statement}

HEACalculator is already positioned as reusable research software rather than a one-off project artifact. The package is archived through Zenodo \parencite{sariturk2019}, distributed as installable Python software, documented for CLI and API use, and supported by continuous integration workflows. Its scientific correctness is also checked directly against the literature: the test suite reproduces phase predictions for 120 reference compositions drawn from the eight source-model papers and preserves close agreement with published continuous parameters.

The software has also been adopted outside its original codebase as a feature-generation tool in subsequent materials informatics studies. Zadeh et al.\ used HEACalculator as part of the descriptor pipeline for a data-driven study of phase compatibility in NiTi shape memory alloys \parencite{zadeh2024}, and Sheikh et al.\ used it to generate domain-specific features for machine-learning prediction of hardness in refractory high-entropy alloys \parencite{sheikh2025}. These examples show that the package is useful not only for direct phase-screening calculations, but also as a reusable source of physically informed alloy descriptors in broader computational materials workflows.

\section{Acknowledgements}

Portions of this research were conducted with the advanced computing resources provided by Texas A\&M High Performance Research Computing. The calculations reported in this paper were partially performed at TUBITAK ULAKBIM, High Performance and Grid Computing Center (TRUBA resources).

\printbibliography

@article{cantor2004,
  author  = {Cantor, B. and Chang, I. T. H. and Knight, P. and Vincent, A. J. B.},
  title   = {Microstructural development in equiatomic multicomponent alloys},
  journal = {Materials Science and Engineering: A},
  year    = {2004},
  volume  = {375--377},
  pages   = {213--218},
  doi     = {10.1016/j.msea.2003.10.257}
}

@article{yeh2004,
  author  = {Yeh, J.-W. and Chen, S.-K. and Lin, S.-J. and Gan, J.-Y. and Chin, T.-S.
             and Shun, T.-T. and Tsau, C.-H. and Chang, S.-Y.},
  title   = {Nanostructured high-entropy alloys with multiple principal elements:
             novel alloy design concepts and outcomes},
  journal = {Advanced Engineering Materials},
  year    = {2004},
  volume  = {6},
  pages   = {299--303},
  doi     = {10.1002/adem.200300567}
}

@article{george2019,
  author  = {George, E. P. and Raabe, D. and Ritchie, R. O.},
  title   = {High-entropy alloys},
  journal = {Nature Reviews Materials},
  year    = {2019},
  volume  = {4},
  pages   = {515--534},
  doi     = {10.1038/s41578-019-0121-4}
}

@article{miracle2017,
  author  = {Miracle, D. B. and Senkov, O. N.},
  title   = {A critical review of high entropy alloys and related concepts},
  journal = {Acta Materialia},
  year    = {2017},
  volume  = {122},
  pages   = {448--511},
  doi     = {10.1016/j.actamat.2016.08.081}
}

@article{yang2012,
  author  = {Yang, X. and Zhang, Y.},
  title   = {Prediction of high-entropy stabilized solid-solution in multi-component alloys},
  journal = {Materials Chemistry and Physics},
  year    = {2012},
  volume  = {132},
  pages   = {233--238},
  doi     = {10.1016/j.matchemphys.2011.11.021}
}

@article{guo2011,
  author  = {Guo, S. and Ng, C. and Lu, J. and Liu, C. T.},
  title   = {Effect of valence electron concentration on stability of fcc or bcc
             phase in high entropy alloys},
  journal = {Journal of Applied Physics},
  year    = {2011},
  volume  = {109},
  pages   = {103505},
  doi     = {10.1063/1.3587228}
}

@article{wang2015,
  author  = {Wang, Z. and Huang, Y. and Yang, Y. and Wang, J. and Liu, C. T.},
  title   = {Atomic-size effect and solid solubility of multicomponent alloys},
  journal = {Scripta Materialia},
  year    = {2015},
  volume  = {94},
  pages   = {28--31},
  doi     = {10.1016/j.scriptamat.2014.09.010}
}

@article{singh2014,
  author  = {Singh, A. K. and Kumar, N. and Dwivedi, A. and Subramaniam, A.},
  title   = {A geometrical parameter for the formation of disordered solid solutions
             in multi-component alloys},
  journal = {Intermetallics},
  year    = {2014},
  volume  = {53},
  pages   = {112--119},
  doi     = {10.1016/j.intermet.2014.04.019}
}

@article{ye2015scr,
  author  = {Ye, Y. F. and Wang, Q. and Lu, J. and Liu, C. T. and Yang, Y.},
  title   = {The generalized thermodynamic rule for phase selection in multicomponent
             alloys},
  journal = {Scripta Materialia},
  year    = {2015},
  volume  = {104},
  pages   = {53--55},
  doi     = {10.1016/j.scriptamat.2015.03.010}
}

@article{troparevsky2015,
  author  = {Troparevsky, M. C. and Morris, J. R. and Kent, P. R. C.
             and Lupini, A. R. and Stocks, G. M.},
  title   = {Criteria for predicting the formation of single-phase high-entropy
             alloys},
  journal = {Physical Review X},
  year    = {2015},
  volume  = {5},
  number  = {1},
  pages   = {011041},
  doi     = {10.1103/PhysRevX.5.011041}
}

@article{senkov2016,
  author  = {Senkov, O. N. and Miracle, D. B.},
  title   = {A new thermodynamic parameter to predict formation of solid solution
             or intermetallic phases in high entropy alloys},
  journal = {Journal of Alloys and Compounds},
  year    = {2016},
  volume  = {658},
  pages   = {603--607},
  doi     = {10.1016/j.jallcom.2015.10.279}
}

@article{king2016,
  author  = {King, D. J. M. and Middleburgh, S. C. and McGregor, A. G.
             and Cortie, M. B.},
  title   = {Predicting the formation and stability of single phase high-entropy
             alloys},
  journal = {Acta Materialia},
  year    = {2016},
  volume  = {104},
  pages   = {172--179},
  doi     = {10.1016/j.actamat.2015.11.051}
}

@article{takeuchi2005,
  author  = {Takeuchi, A. and Inoue, A.},
  title   = {Classification of bulk metallic glasses by atomic size difference,
             heat of mixing and period of constituent elements and its application
             to characterization of the main alloying element},
  journal = {Materials Transactions},
  year    = {2005},
  volume  = {46},
  pages   = {2817--2829},
  doi     = {10.2320/matertrans.46.2817}
}

@article{martin2022heaps,
  author  = {Martin, P. and Madrid-Cortes, C. E. and C{\'a}ceres, C.
             and Araya, N. and Aguilar, C. and Cabrera, J. M.},
  title   = {{HEAPS}: A user-friendly tool for the design and exploration
             of high-entropy alloys based on semi-empirical parameters},
  journal = {Computer Physics Communications},
  year    = {2022},
  volume  = {278},
  pages   = {108398},
  doi     = {10.1016/j.cpc.2022.108398},
  url     = {https://doi.org/10.1016/j.cpc.2022.108398}
}

@article{ward2018matminer,
  author  = {Ward, L. and Dunn, A. and Faghaninia, A. and Zimmermann, N. E. R.
             and Bajaj, S. and Wang, Q. and Montoya, J. H. and Chen, J.
             and Bystrom, K. and Dylla, M. and Chard, K. and Asta, M.
             and Persson, K. A. and Snyder, G. J. and Foster, I. and Jain, A.},
  title   = {Matminer: An open source toolkit for materials data mining},
  journal = {Computational Materials Science},
  year    = {2018},
  volume  = {152},
  pages   = {60--69},
  doi     = {10.1016/j.commatsci.2018.05.018}
}

@article{subedi2021pympealab,
  author  = {Subedi, U. and Kunwar, A. and Coutinho, Y. A. and Gyanwali, K.},
  title   = {{pyMPEALab} toolkit for accelerating phase design in
             multi-principal element alloys},
  journal = {Metals and Materials International},
  year    = {2021},
  volume  = {27},
  pages   = {2157--2168},
  doi     = {10.1007/s12540-021-01100-9}
}

@article{subedi2022,
  author  = {Subedi, U. and Coutinho, Y. A. and Malla, P. B.
             and Gyanwali, K. and Kunwar, A.},
  title   = {Automatic featurization aided data-driven method for estimating
             the presence of intermetallic phase in multi-principal element alloys},
  journal = {Metals},
  year    = {2022},
  volume  = {12},
  pages   = {964},
  doi     = {10.3390/met12060964}
}

@misc{sariturk2019,
  author    = {Sarıtürk, Doğuhan},
  title     = {{HEACalculator}},
  year      = {2019},
  publisher = {Zenodo},
  doi       = {10.5281/zenodo.3590318},
  url       = {https://doi.org/10.5281/zenodo.3590318}
}

@article{zadeh2024,
  author  = {Zadeh, Sina Hossein and Cakirhan, Cem and Khatamsaz, Danial
             and Broucek, John and Brown, Timothy D. and Qian, Xiaoning
             and Karaman, Ibrahim and Arroyave, Raymundo},
  title   = {Data-driven study of composition-dependent phase compatibility
             in {NiTi} shape memory alloys},
  journal = {Materials and Design},
  year    = {2024},
  volume  = {244},
  pages   = {113096},
  doi     = {10.1016/j.matdes.2024.113096},
  url     = {https://doi.org/10.1016/j.matdes.2024.113096}
}

@article{sheikh2025,
  author  = {Sheikh, Sofia and Zadeh, Sina Hossein and Mulukutla, Mrinalini
             and Hastings, Trevor and Arroyave, Raymundo},
  title   = {Predicting hardness in refractory high-entropy alloys using
             machine learning},
  journal = {Materials Letters},
  year    = {2025},
  volume  = {400},
  pages   = {138940},
  doi     = {10.1016/j.matlet.2025.138940},
  url     = {https://doi.org/10.1016/j.matlet.2025.138940}
}

@article{guo2013,
  author  = {Guo, S. and Hu, Q. and Ng, C. and Liu, C. T.},
  title   = {More than entropy in high-entropy alloys: Forming solid solutions
             or amorphous phase},
  journal = {Intermetallics},
  year    = {2013},
  volume  = {41},
  pages   = {96--103},
  doi     = {10.1016/j.intermet.2013.05.005}
}

\end{document}